\documentstyle[12pt]{article}

\def\hybrid{\topmargin -20pt    \oddsidemargin 0pt
        \headheight 0pt \headsep 0pt
        \textwidth 6.25in       
        \textheight 9.5in       
        \marginparwidth .875in
        \parskip 5pt plus 1pt   \jot = 1.5ex}

\def\ket#1{|{#1}\rangle}
\def\noi{\noindent}

\def\baselinestretch{1.2}

\catcode`\@=11

\def\marginnote#1{}
\def\draftlabel#1{{\@bsphack\if@filesw {\let\thepage\relax
   \xdef\@gtempa{\write\@auxout{\string
      \newlabel{#1}{{\@currentlabel}{\thepage}}}}}\@gtempa
   \if@nobreak \ifvmode\nobreak\fi\fi\fi\@esphack}
        \gdef\@eqnlabel{#1}}
\def\@eqnlabel{}
\def\@vacuum{}
\def\draftmarginnote#1{\marginpar{\raggedright\scriptsize\tt#1}}

\def\draft{\oddsidemargin -.2truein
        \def\@oddfoot{\sl preliminary draft \hfil
        \rm\thepage\hfil\sl\today\quad\militarytime}
        \let\@evenfoot\@oddfoot \overfullrule 3pt
        \let\label=\draftlabel
        \let\marginnote=\draftmarginnote
   \def\@eqnnum{(\theequation)\rlap{\kern\marginparsep\tt\@eqnlabel}%
\global\let\@eqnlabel\@vacuum}  }

\def\preprint{\twocolumn\sloppy\flushbottom\parindent 2em
        \leftmargini 2em\leftmarginv .5em\leftmarginvi .5em
        \oddsidemargin -.5in    \evensidemargin -.5in
        \columnsep .4in \footheight 0pt
        \textwidth 10.in        \topmargin  -.4in
        \headheight 12pt \topskip .4in
        \textheight 6.9in \footskip 0pt
        \def\@oddhead{\thepage\hfil\addtocounter{page}{1}\thepage}
        \let\@evenhead\@oddhead \def\@oddfoot{} \def\@evenfoot{} }

\def\numberbysection{\@addtoreset{equation}{section}
        \def\theequation{\thesection.\arabic{equation}}}

\def\underline#1{\relax\ifmmode\@@underline#1\else
        $\@@underline{\hbox{#1}}$\relax\fi}

\def\titlepage{\@restonecolfalse\if@twocolumn\@restonecoltrue
\onecolumn
     \else \newpage \fi \thispagestyle{empty}\c@page\z@
        \def\thefootnote{\fnsymbol{footnote}} }

\def\endtitlepage{\if@restonecol\twocolumn \else \newpage \fi
        \def\thefootnote{\arabic{footnote}}
        \setcounter{footnote}{0}}  

\catcode`@=12
\relax

\def\figcap{\section*{Figure Captions\markboth
        {FIGURECAPTIONS}{FIGURECAPTIONS}}\list
        {Figure \arabic{enumi}:\hfill}{\settowidth\labelwidth{Figure
999:}
        \leftmargin\labelwidth
        \advance\leftmargin\labelsep\usecounter{enumi}}}
 \relax
\def\tablecap{\section*{Table Captions\markboth
        {TABLECAPTIONS}{TABLECAPTIONS}}\list
        {Table \arabic{enumi}:\hfill}{\settowidth\labelwidth{Table
999:}
        \leftmargin\labelwidth
        \advance\leftmargin\labelsep\usecounter{enumi}}}
 \relax
\def\reflist{\section*{References\markboth
        {REFLIST}{REFLIST}}\list
        {[\arabic{enumi}]\hfill}{\settowidth\labelwidth{[999]}
        \leftmargin\labelwidth
        \advance\leftmargin\labelsep\usecounter{enumi}}}
 \relax
\makeatletter
\newcounter{pubctr}
\def\publist{\@ifnextchar[{\@publist}{\@@publist}}
\def\@publist[#1]{\list
        {[\arabic{pubctr}]\hfill}{\settowidth\labelwidth{[999]}
        \leftmargin\labelwidth
        \advance\leftmargin\labelsep
        \@nmbrlisttrue\def\@listctr{pubctr}
        \setcounter{pubctr}{#1}\addtocounter{pubctr}{-1}}}
\def\@@publist{\list
        {[\arabic{pubctr}]\hfill}{\settowidth\labelwidth{[999]}
        \leftmargin\labelwidth
        \advance\leftmargin\labelsep
        \@nmbrlisttrue\def\@listctr{pubctr}}}
 \relax
\makeatother
\newskip\humongous \humongous=0pt plus 1000pt minus 1000pt

\newif\ifdtup

\font\Scbig=cmss10 scaled\magstep1
\font\Scscr=cmss8 scaled\magstep1
\font\Scscrscr=cmss8
\newfam\Scfam
\textfont\Scfam=\Scbig
\scriptfont\Scfam=\Scscr
\scriptscriptfont\Scfam=\Scscrscr

\relax
\hybrid
\def\lvm{\leavevmode\hbox to\parindent{\hfill}}

\def\thefootnote{\fnsymbol{footnote}}
\def\BE{\begin{equation}}
\def\EE{\end{equation}}
\def\BA{\begin{eqnarray}}
\def\EA{\end{eqnarray}}

\def\tt{\bar\tau}

\def\lvm{\leavevmode\hbox to\parindent{\hfill}}
\def\bar{\overline}

\def\BE{\begin{equation}}
\def\EE{\end{equation} \vskip 0.30\baselineskip}
\def\BA{\begin{array}}
\def\EA{\end{array}}

\def\noi{\noindent}

\def\frac#1#2{{\textstyle{{#1}\over{#2}}}}

\def\ket#1{|{#1}\rangle}

\newif\ifold \oldtrue 
\let\ssection=\section
\def\section{\setcounter{equation}{0}\ssection}

\begin{document}
\renewcommand{\theequation}{\arabic{equation}}
\newcommand{\beq}{\begin{equation}}
\newcommand{\eeq}[1]{\label{#1}\end{equation}}
\newcommand{\ber}{\begin{eqnarray}}
\newcommand{\eer}[1]{\label{#1}\end{eqnarray}}
\begin{titlepage}
\begin{center}

\hfill IMAFF-FM-02/14\\
\hfill hep-th/0212348
\vskip .4in

{\large \bf Some Issues in Conformal Field Theory
with Boundaries and Crosscaps}
\vskip 2in

{\bf Beatriz Gato-Rivera}
\vskip .3in

 {\em Instituto de Matem\'aticas y F\'\i sica Fundamental, CSIC,\\
 Serrano 123, Madrid 28006, Spain} \footnote{Invited talk given at
the Conference `Conformal Field Theory and Integrable Systems',
held by Landau Institute, Russia, September 15-21, 2002.
e-mail address: bgator@imaff.cfmac.csic.es}\\
\vskip .2in

{\em NIKHEF-H, Kruislaan 409, NL-1098 SJ Amsterdam, The Netherlands}\\

\vskip 1.0in

\end{center}

\begin{center} {\bf ABSTRACT } \end{center}
\begin{quotation}
This is a brief introduction to the subject of Conformal Field Theory
on surfaces with boundaries and crosscaps, which 
describes the perturbative expansion of open string theory.

\end{quotation}
\vskip 1.5cm

December 2002 
\end{titlepage}

\def\baselinestretch{1.2}
\baselineskip 17 pt

I am going to give a brief introduction to the subject of
CFT on surfaces with boundaries and crosscaps, which
is the CFT description of perturbative open string theory.
Let me start with a small historical remark. During many years
open string theory has been much less popular than closed string
theory and there were reasons for it. On the one hand, it was
believed that open strings had nothing to do with the real world
since they could not provide realistic models for particle Physics,
in contrast to the case of the heterotic strings. On the other hand,
open strings are much more difficult to deal technically than 
closed strings. For these reasons it is understandable that,
from the `rise' of String Theory in 1985 until 1995, the bulk of
string theoretists concentrated on closed strings, mainly
heterotic strings, while only very few paid attention to open 
strings \cite{OPSPS} \cite{RomeP0}.
This situation changed drastically when Polchinski discovered
D-branes in 1995 \cite{Pol}. The reason is that, apart from
the genuine interest on these new objects to which the open
strings attach, it was also shown that the open strings are dual
to closed strings, in particular to the heterotic strings, and finally
it appeared that all known string theories are related to each other
(as a matter of fact, as everybody knows by now,
it seems that they are different limits of one
and the same theory in 11 dimensions, known as M-theory).
This was the end of the `discrimination' against open string theory
which at present is a very active field. However, the CFT treatment
of open strings is been worked out by very few groups and this is
the reason I decided to give an introductory talk in this conference.

In what follows I will describe the basic ingredients necessary for 
the study of CFT's on surfaces with boundaries and crosscaps. Let
us first review the 2-dimensional surfaces involved in the perturbative
expansion of string theories at one-loop, that is the worldsheets swept 
by the strings when moving in space-time. In string theory at one-loop
one finds four different types of topologically inequivalent worldsheets:
the torus, the Klein bottle, the annulus, and the M\"obius strip. Depending
on the type of string theory, some of these surfaces may or may not 
appear. For closed oriented strings only the torus appear whereas for 
closed unoriented strings one has also the Klein bottle. For open oriented
strings one finds the annulus plus the torus corresponding to the closed
sector of the theory and, finally, for open unoriented strings the four 
types of worldsheets appear. The `direct channel' surfaces (with the
proper time making a loop) contain the information
about the complete spectrum of the CFT defined on them; that is, the
amplitudes of these surfaces represent the partition functions of the
theory that give the number of states level by level. But one can also
look at these surfaces from the `transverse channel' point of view,
that is exchanging the space and time coordinates of the worldsheet.
The torus does not change, but the annulus, the M\"obius strip and
the Klein bottle look quite different. The annulus is converted into a
cylinder between two boundaries, that shows the propagation of
closed strings between the two boundaries, the M\"obius strip is
converted into a cylinder between a boundary and a crosscap
(the M\"obius strip has only one boundary), and the Klein bottle turns 
into a cylinder between two crosscaps (a crosscap is a boundary 
with the opposite sides identified).

Now let us consider any 2-dimensional surface with boundaries 
and crosscaps. Let us call the closed string chiral algebra as
$G_L \otimes G_R$, with generators $W_n$ and $\hat W_n$. At a
boundary or crosscap only the diagonal combination survives
because these `defects' interchange left and right-movers. As a
result, the boundary states $\ket{B}$ and the crosscap states  
$\ket{C}$ must satisfy the conditions

\BE
[W_n - (-1)^{h_w} \ w \ \hat W_{-n}] \ket{B} = 0, \qquad
[W_n - (-1)^{h_w + n} \ w \ \hat W_{-n}] \ket{C} = 0,
\EE

\noi
where $h_w$ is the conformal weight of the generator $W_n$
and if $w \neq 1$ this is called a symmetry breaking boundary 
condition \cite{FS1}\cite{SR}. There is a formal solution to
these conditions, given by the so-called Ishibashi states
\cite{Ishi}

\BE
\ket{i}_B = \sum_l \ket{i,l}_L \otimes  U_B \ket{i,l}_R, \qquad
\ket{i}_C = \sum_l \ket{i,l}_L \otimes  U_C \ket{i,l}_R,
\EE

\noi
where $i$ denotes the ground state, $l$ the level, and 
$U_B$ and $U_C$ are anti-unitary operators. The physical
boundary states and crosscap states are linear combinations
of the Ishibashi states

\BE
\ket{B_a} = \sum_i B_{ai} \ket{i}_B, \qquad
\ket{C} = \sum_i \Gamma_{i} \ket{i}_C
\EE

\noi
where $B_{ai}$ and $\Gamma_i$ are called boundary and crosscap
coefficients, respectively. The label $a$ indicates that a CFT can
have different sets of boundaries, whereas it can have only one type
of crosscap.

The boundary and crosscap coefficients are very important quantities
in open string theory since they contain information about the spectrum
of the string states as well as information about the D-branes and 
orientifold planes (O-planes for short). These coefficients are
constrained by sewing constraints, which are rather difficult to solve,
and also by integrality and positivity conditions that are very restrictive
but much easier to solve than the sewing constraints. The origin of the 
integrality and positivity conditions is the fact that the partition functions,
which give the state multiplicities, are given by the one-loop amplitudes
and, in the transverse channel, these amplitudes depend explicitely on
the boundary and crosscap coefficients. For example, from the annulus,
where open strings are running in the direct channel, one gets the
integrand

\BE
N^a \  N^b \  A^i_{ab} \ \chi^i(\tau/2) 
\EE

\noi
where $N^a, \ N^b $ are the Chan-Paton factors at the ends of the strings,
$\chi^i$ is the character of the representation $i$, and $A^i_{ab}$ can be 
obtained from the transverse channel, having the expression

\BE
A^i_{ab} = \sum_j S^i_j \ B_{ja} \ B_{jb}, 
\EE

\noi
where $S$ is the modular matrix, $S: \tau \rightarrow -1/{\tau}$. Now from the
direct channel one gets that

\BE
A^i_{ab} \in {\bf Z} \geq 0.
\EE

\noi
In the same manner, from the M\"obius strip and the Klein bottle one
obtains the conditions

\BE
1/2 (A^i_{aa} + M^i_a) \in  {\bf Z} \geq 0, \qquad 
1/2 (Z_{ii} + K^i) \in  {\bf Z} \geq 0,
\EE

\noi
with 

\BE
 M^i_a = \sum_j P^i_j \ B_{ja} \ \Gamma_j \in  {\bf Z}, \qquad
 K^i = \sum_j S^i_j \ \Gamma_j \ \Gamma_j \in  {\bf Z} ,
\EE

\noi 
where the P matrix is defined as \cite{RomeP0}\cite{RomeP1} 
$P = \sqrt{T} S T^2 S \sqrt{T}$, $T$ being the modular matrix,
$T: \tau \rightarrow \tau + 1$, and $Z_{ii}$
comes from the torus partition function.
(Let us remind that the torus partition function can be expressed as
 $ \sum_{ij} \chi_i(\tau)\ Z_{ij}\ \chi_j(\tau)$).

Another important consistency condition is the tadpole cancellation.
It happens that an open unoriented string with arbitrary Chan-Paton
factors is in general inconsistent due to infrared divergences in
one-loop amplitudes. This is very easy to see in the transverse channel
where, by factorization, the `cylinder' decomposes as the product of
the propagator times the tadpoles corresponding to the extremes
(boundary tadpoles and/or crosscap tadpoles). Therefore the
tadpoles must cancel, what implies that the Chan-Paton factors
must be adjusted to some specific values. But the Chan-Paton
factors reflect the gauge group of the theory and therefore the
tadpole cancellation fixes the possible allowed gauge groups.

Now let us see some examples of solutions for the boundary and
crosscap coefficients. For any CFT such that the modular invariant is
the charge conjugation, that is $Z_{ij} = C_{ij}$, Cardy found 
the boundary coefficients \cite{Cardy}

\BE
B_{ai} = {S_{ai} \over \sqrt{S_{0i}}}
\EE

 \noi
and the `Rome group' (Sagnotti and collaborators) found the
crosscap coefficients \cite{RomeP1}\cite{Rome}

\BE
\Gamma_{i} = {P_{0i} \over \sqrt{S_{0i}}}
\EE

These results have been generalized first of all by allowing
more general simple current generated Klein bottles, and
secondly to arbitrary simple current modular invariants, as
classified in refs. \cite{G-RS} and \cite{KS}. This generalization
was pioneered by the Rome group, who worked out the first examples
for SU(2) \cite{RomeP1}, 
and was completed in a series of papers by Fuchs and
Schweigert \cite{FS2} and by a group from Amsterdam (Huiszoon,
Schellekens and Sousa) \cite{HSS}. The final result can be summarized
very concisely in a single formula for the boundary and crosscap
coefficients \cite{FHSSW}. The simplest example, which corresponds to a
non-standard Klein bottle choice specified by a simple current $k$,
reads

\BE
B^k_{ai} = {S_{ai} \over \sqrt{S_{ki}}}, \qquad
\Gamma^k_{i} = {P_{ki} \over \sqrt{S_{ki}}},
\EE

\noi
where the index $k$ refers to the Klein bottle simple current.

To finish let me say a few words about the meaning of boundaries and
crosscaps in space-time. The boundaries are glued to D-branes as
it is easy to see intuitively in the open string picture where the ends
of the strings are attached to the D-brane. Crosscaps, however, are
very difficult to visualize since, first of all, they are not even localized.
This means that while one can `see' a boundary, even touch it with
the fingers on an annulus or a M\"obius strip, one cannot do the 
same with a crosscap. In space-time crosscaps are related 
to orientifold configurations in such a way that a single
crosscap can correspond to a sum of orientifolds, not to just one.
(Orientifolds are static hyperplanes where left and right-movers are
identified, they are like mirrors for charge conjugation of D-branes).   
 
\vskip .7in
\centerline{\bf Acknowledgements}

I am very grateful to A. Sagnotti and A.N. Schellekens for getting me started
on the subject of CFT with boundaries and crosscaps.

\end{document}